\documentclass[prb,twocolumn,amsmath,amssymb,showpacs]{revtex4}

\usepackage{graphicx}
\usepackage{epstopdf}
\usepackage{bm}
\usepackage{float}

\begin{document}

\title{Counting the Number of Excited States in Organic Semiconductor Systems
Using Topology}

\author{Michael J. \surname{Catanzaro}$^a$} \author{Tian \surname{Shi}$^b$}
\author {Sergei \surname{Tretiak}$^c$} \email{serg@lanl.gov} \author {Vladimir
  Y. \surname{Chernyak}$^{a,b}$} \email{chernyak@chem.wayne.edu}
  \affiliation{$^a$Department of Mathematics, Wayne State University, 656 W.
  Kirby, Detroit, MI 48202} \affiliation{$^b$Department of Chemistry, Wayne
  State University, 5101 Cass Ave, Detroit, MI 48202}
  \affiliation{$^c$Theoretical Division, Center for Nonlinear Studies,  and
  Center for Integrated Nanotechnologies, Los Alamos National Laboratory,  Los
Alamos, NM 87545} \date{December 22, 2014}

\begin{abstract}

Exciton Scattering (ES) theory attributes excited electronic states to standing
waves in quasi-one-dimensional molecular materials by assuming a quasi-particle
picture of optical excitations. The quasi-particle properties at branching
centers are described by the corresponding scattering matrices.  Here we
identify the topological invariant of a scattering center, referred to as its
winding number, and apply topological intersection theory to count the number
of quantum states in a quasi-one-dimensional system.  \end{abstract}

\maketitle

\section{Introduction}
\label{sec:Intro}

A quantum system that resides on a network/graph forms an important
generalization of a particle in a box problem, where the quantum eigenstates
can be attributed to standing waves. In our earlier work, we have developed a
multi-scale Exciton Scattering (ES) approach for photo-excitations in large,
conjugated molecules. Excitations are described as quasi-particles (excitons)
that move along the linear segments, getting scattered at molecular termini,
joints, and branching centers, represented by graph edges and vertices,
respectively~\cite{ES-naturephys,ES1,ES2,ES3,ES-prl}. Excited states can
then be calculated by solving the ES equations [see Eq.~(\ref{ES-equations})],
once the ES parameters, namely exciton dispersion $\omega(k)$ for graph edges
and scattering matrices $\Gamma_{a}(k)$ for graph vertices $a$, are known. Both
ingredients $\omega(k)$ and $\Gamma_{a}(k)$ can be extracted from the
reference quantum chemistry calculations of relatively small molecular
fragments with reasonably low computational cost~\cite{ES2,ES-prl}. Thus,
extracting the ES parameters and solving the ES equations are two
fundamental steps in the multi-scale modeling of electronic excitations in
branched, organic semiconductors. The ES approach is asymptotically exact for
long enough linear segments~\cite{ES1,ES-prl} and shows excellent agreement
with reference quantum chemistry calculations~\cite{ES3,ES-prl} even in the
case of short linear segments. Our previous studies have demonstrated accurate
calculations of excited state transition energies~\cite{ES3,ES-prl} and optical
spectra~\cite{ES-dipole} in very large, low-dimensional molecular systems with
negligible computational cost using this technique. Moreover, this framework
enables an efficient characterization of excited-state electronic structure
modifications due to different branching pattern and donor/acceptor
substitutions~\cite{ES-DA,ES-analytical}. We reiterate that all information on
the system Hamiltonian and further approximations (e.g., the quantum chemistry
method involved, basis sets, etc.) is fully contained in the tabulated ES
parameters [$\omega(k)$ and $\Gamma_{a}(k)$], so once they are extracted, there
is no need to specify the aforementioned Hamiltonian, as well as supporting
information.

It is worth noting that the ES approach expresses electronic excitations in
branched, conjugated molecules in terms of excited states in infinite polymers
with perfect geometry, and the scattering properties of molecular joints. Due to
discrete translational symmetry, electronic excitations in an infinite polymer
chain possess a good quantum number, referred to as quasimomentum $k \in
[-\pi,\pi]$, that resides in a $1$ dimensional Brillouin zone. This Brillouin zone is
represented by a circle, obtained from the segment $[-\pi,\pi]$ by implementing
periodic boundary conditions. The continuous spectrum of electronic excitations
in an infinite polymer chain is, therefore, represented by exciton bands, each
characterized by its dispersion $\omega(k)$, i.e., the dependence of the
excitation energy $\omega$ on its quasimomentum $k$. The ES methodology
implements a well-known particle in a box concept by viewing a branched,
conjugated molecule as a quasi-one-dimensional box, where the linear segments
are represented by sub-boxes, while the scattering matrices $\Gamma_{a}(k)$
play the role of the proper boundary conditions obtained from this microscopic
approach. In a finite branched structure, standing waves are formed due to
interference of waves from quasi-particle (exciton) scattering at molecular
joints and termini. This results in a discrete spectrum of electronic
excitations, in full analogy with the standard particle in a box problem. The
exciton energies for a finite molecule are obtained by solving a generalized
spectral problem in a $n= 2n_{{\rm s}}$-dimensional vector space, with $n_{{\rm
s}}$ being the number of linear segments (graph edges), referred to as the
system of the ES equations~\cite{ES1}. In general terms, the ES approach splits
the energy range into spectral regions, spanned by different exciton bands, and
further identifies the exciton spectra within the above regions by solving the
ES equations.  Obviously the exciton bands are properties of infinite
polymers, or stated differently, properties of the repeat unit, whereas
the excitation energies also depend on the scattering matrices of the
molecular joints, and the molecular structure, i.e. the linear segment lengths
$L_{\alpha}$ and the way the segments are connected via the joints.

The spectrum of electronic excitations is a key fingerprint that defines
molecular photoinduced dynamics, including energy transfer and relaxation
processes. Naturally, an important property of the excited state manifold
determined by a specific exciton band is the total number $N$ of excited states
in the given spectral region. According to the ES theory, $N$ is determined by
the spectrum $\omega(k)$ of the relevant band, the scattering matrices
$\Gamma_{a}(k)$ of the molecular joints and termini, the linear segment lengths
$L_{\alpha}$ (that can be measured as the numbers of repeat units within a
segment), and finally the molecular topology (i.e., the way the linear segments
are connected). The molecular topology can be formally described by a graph,
whose vertices and edges represent the joints (including termini) and linear
segments respectively. Topologically speaking, a joint $a$ is
determined by its degree $r_{a}$.  Equivalently, all termini (degree one), all double
joints (degree two), all triple joints (degree three) and so on, are
topologically identical; the differences are encoded in their scattering
matrices. A natural question arises: Is there a simple relation that expresses
the number $N$ of excitations within the spectral region of a given exciton
band in terms of the inputs of the ES approach, i.e., the graph that describes the
topological structure of a branched molecule, the dispersion $\omega(k)$,
scattering matrices $\Gamma_{a}(k)$ and the segment lengths $L_{\alpha}$, where
$a$ and $\alpha$ stand for the graph vertices and edges, respectively?

In our previous work, we made the first attempt to address the above question
within the ES approach. We introduced a topological invariant (coining
winding number) to count the total number $N$ of excited states within the
spectral region of a single exciton band in symmetric, conjugated
molecules~\cite{ES-symmetric}. The counting was achieved by associating to
a scattering matrix $\Gamma_{a}(k)$ of a highly symmetric molecular scattering
center, an integer-valued topological invariant $Q_{a}$, coined topological
charge. Further, we implemented topological intersection theory in its very
elementary, albeit very intuitive form, resulting in a lower bound
\begin{eqnarray}
\label{bound-symmetric} N \ge \sum_{\alpha}L_{\alpha}+ \sum_{a}Q_{a}.
\end{eqnarray}
We have also demonstrated that for molecules with long enough linear segments,
the bound in Eq.~(\ref{bound-symmetric}) becomes tight, i.e., the inequality
turns into an equality, so that the count is solely dependent on the number of
repeat units and the scattering matrices of joints. The obtained result allows
for a natural interpretation: In a finite-size molecule, the number of
electronic excitations within the spectral region of the exciton band is given
by the total number of repeat units in the linear segments plus the sum of
topological charges of the scattering centers. In other words, each repeat unit provides
a state to the full count, whereas the number of states provided by a
scattering center (a molecular terminal or a joint) are defined by its
topological charge $Q_{a}$. The latter is fully determined by the topological
properties of the relevant scattering matrix $\Gamma_{a}(k)$. We applied this
method to investigate the scattering matrix of the 'X' joint of phenylacetylene
(PA) based molecules, and successfully predicted one additional resonance state
with energy inside the band and two bound excited states with energy outside
the band contributed from the joint~\cite{ES-symmetric}.

This methodology has already found applications in building effective tight
binding models for electronic excitations in low-dimensional conjugated
systems~\cite{ES-tight-binding}. Since the number of lattice sites for a
scattering center/repeat unit in these models depends explicitly on the
topological invariant, we have a systematic way to construct the lattice model,
and therefore, the Frenkel type exciton Hamiltonian. For example, every repeat
unit of PA based molecules can naturally be represented by a lattice site,
since each of them contributes exactly one excited state to an exciton band.
Another non-trivial example arises from scattering at $ortho$ and $meta$ joints
contributing one and zero additional states, translating into one and zero
lattice sites for them, respectively. The number of lattice sites for the 'X'
joint is one, since it provides one additional state. Such assignment of
lattice sites guarantees the correctness and accuracy of the tight binding
models.

It is important to note that the results on the counting, discussed above, were
ultimately dependent on the high symmetry of the molecules under study. Not only
should the joints possess high symmetry, but the latter should also be
preserved by the actual molecule. Stated differently, the aforementioned
results are valid only for linear oligomers and for molecules where a
symmetrical double, triple, or quadruple joint connects linear segments of equal
length. The latter condition is necessary for the approach we used in
order to maintain the high symmetry of the system. This limitation is imposed
by restrictive assumptions: The main result (Eq.~(\ref{bound-symmetric})), has
actually been derived for linear oligomers, followed by extending it to highly
symmetric molecules, listed above, using the classification of electronic
excitations by their symmetry, thus mapping the counting problem in a symmetric
molecule to its linear segment counterpart. Therefore, the approach presented
in~\cite{ES-symmetric} is not extendable to an arbitrary molecular topology.
Moreover, it provides no insights on how the notion of topological charge could
be extended to the case of an arbitrary joint, or if such extension is possible at
all.

In this manuscript, we address the question of counting electronic excitations
within the spectral region of a single exciton band, explicitly formulated
earlier, in full generality, i.e., for a branched molecule
described by an arbitrary graph, as well as general type scattering centers,
not necessarily possessing a high degree of symmetry.  Specifically, we
introduce a topological invariant by identifying the winding number for an {\em
arbitrary} scattering center and relate the segment lengths $L_{\alpha}$ along
with the winding numbers $l_a$ of the vertices to the total number of excited
states $N$ in a single exciton band. We obtain an explicit lower bound for $N$
in terms of $L_{\alpha}$ and $l_a$ [Eqs.~(\ref{number-of-excitons}) and
(\ref{bound-explicit})]. This is achieved by formulating the ES equations as
an intersection problem, followed by applying topological intersection
theory, which provides simple expressions for the {\it intersection index},
the latter bounding the number of solutions of the ES equations. Furthermore,
this bound becomes exact when the molecular arms are long enough. We also
demonstrate the formulated concepts by analyzing excited state structure in
branched conjugated polyfluorenes. We emphasize that the theory presented
here applies generally to any quasi-one-dimensional quantum system, as our
results are based on the following generic properties: (i) unitarity of
scattering processes, (ii) phase change from propagation along linear
segments, and (iii) quasi-momenta residing in a one-dimensional Brillouin
zone.

The manuscript is organized as follows. In Section~\ref{sec:Method-Satement},
we introduce the approach we have taken, including a simple and intuitive
description of topological intersection theory, with an emphasis on its
application to the problem under study (Section~\ref{sec:top-intersection}). We
also describe in some detail the main results presented in the manuscript
(Section~\ref{sec:ES-eq-main-results}), as well as intuitive arguments that
stand behind the formal derivations. We present a simple picture of the 
winding number (Section~\ref{sec:winding-index}), together with intuitive
arguments in support of the Index Theorem. Moreover,
we rationalize a relation between solutions of the ES equation
and actual excitons (Section~\ref{sec:ES-solution-to-exciton}).
The reader not interested in the details
of these derivations can then bypass this section, jumping directly to
Section~\ref{sec:Application}, where the applications are presented.
Section~\ref{sec:Derivation-details} contains a detailed derivation of our main
results, consisting of several subsections, and focuses on specific steps of
the derivation. In subsection~\ref{sec:ESeqns}, we rewrite the ES equations as
an eigenvalue problem [Eq.~(\ref{ES-equations-vector})], simplifying their
description and allowing for a clear formulation in terms of intersection
theory. Using this representation, we relate the number of solutions to the
number of excitons in Eq.~(\ref{number-of-excitons}). In
subsection~\ref{sec:index-theorem}, we introduce a topological invariant of a
scattering vertex, as well as the local intersection indices. This allows us to
relate the number of solutions to local intersection indices via the index
theorem [Eq.~(\ref{index-theorem})]. We then formulate counting of excitons in
terms of intersection theory in subsection~\ref{sec:sketch-proof-index}, along
with sketching a proof of the index theorem. In subsection~\ref{sec:Long-arm},
we apply perturbation theory to determine the situations in which our
approximation becomes exact. Finally, in Section~\ref{sec:Application}, we
exemplify the presented theory to a class of polyflourene-based molecules. A
brief summary of the results, as well as a list of problems to be addressed in
the future are provided in Section~\ref{sec:Conclusion}.

\section{Methodology and Statements of the Results}
\label{sec:Method-Satement}

In this section, we present our main results on counting the number of
electronic excitations in a branched, conjugated molecule within the spectral
region of a single exciton band, and introduce the key structures involved in
the main statements. We further present clear, intuitive arguments that stand
behind the formal derivations, doing so on a conceptual level.  In this way, a
reader who is interested in the results themselves, as well as the concepts
that stand behind them, can skip the forthcoming
Section~\ref{sec:Derivation-details}, where the details of the derivation are
presented, and jump directly to Section~\ref{sec:Application}, where an
application of our approach is presented. We also introduce the essential concepts
behind topological intersection theory, and how it can be used to solve
systems of equations. This is done by providing a simple example of
intersection theory in terms of single variable calculus, followed by showing
how it can be adapted to solve the ES equations.


\subsection{Exciton Scattering (ES) equations and the main results}
\label{sec:ES-eq-main-results}

Within the ES approach, a branched conjugated molecule is described by the
following data. (i) The first ingredient is the graph that determines the
molecule's topological structure. Generally, a graph is a collection of vertices
connected to one another by edges. Our graph models the conjugated molecule
under study, by assigning a vertex to each scattering center, and an edge to
each molecular linear segment. The graph is equipped with the following
additional data: (ii) the number of repeat units $L_{\alpha}$ in a linear
segment $\alpha$, with $\alpha= 1,\ldots, n_{{\rm s}}$, often referred to as
the segment lengths, (iii) the exciton spectrum (or, equivalently, dispersion)
$\omega(k)$, i.e., the dependence of the exciton energy $\omega$ on its
momentum $k$ in an infinitely long polymer with the same repeat unit, and (iv)
the quasimomentum-dependent exciton scattering matrix $\Gamma_{a}(k)$ of vertex
$a$, where $a$ varies over teh set of molecule scattering centers (including the
termini).


The main exciton counting result is formulated in terms of the winding numbers
$l_{a}$ associated with the molecular scattering centers (graph vertices)
labeled by $a$ and two additional integers $d_{k}$, for $k=0,\pi$, referred to
as bandedge indices. A vertex winding number $l_{a}$ is an integer number
completely determined by the scattering matrix $\Gamma_{a}(k)$ of a vertex $a$.
The bandedge indices $d_{k}$ are integer numbers
that reflect the properties of the bandedge (i.e., $k= 0,\pi$) solutions of the
ES equations, i.e. the relation between the number of solutions and the actual
bandedge excitons. The quantities $l_{a}$ and $d_{k}$ will be introduced later
in this section.

The main counting result is the lower bound
\begin{eqnarray}
\label{exciton-bound-general} N \ge \sum_{\alpha}L_{\alpha}+ \frac{1}{2}\sum_{a}l_{a}+ \frac{d_{0}+ d_{\pi}}{2}
\end{eqnarray}
for the number $N$ of electronic excitations within the spectral region of a
single exciton band, combined with the statement that the bound becomes tight
(i.e., the inequality turns into an equality) for a molecule with long enough
linear segments. There is a very important particular case, which turns out to
represent the generic situation: for all molecular vertices whose scattering
matrices we have extracted from numerical data in our previous studies, as well
as for all scattering matrices calculated analytically within the framework of
lattice models (except for some special values of there parameters), we
find $\Gamma_{a}(0)= \Gamma_{a}(\pi)= -1$. In this case, as will be shown
below, we have $d_{0}= d_{\pi}= -n_{{\rm s}}$; therefore, we introduce the
{\it topological charge} of any vertex $a$ as
\begin{eqnarray}
\label{define-Q-a} Q_{a}= \frac{l_{a}- r_{a}}{2}
\end{eqnarray}
(we reiterate that $r_{a}$ is the vertex degree, i.e., the number of molecule
linear segments attached to the scattering center).
We can further make use of an obvious relation $\sum_{a}r_{a}= 2n_{{\rm s}}$,
which is true in an graph, and recast
Eq.~(\ref{exciton-bound-general}) in the form of Eq.~(\ref{bound-symmetric}).
Hence, we can interpret the topological charge $Q_{a}$ of a vertex as the
number of additional states the corresponding scattering center brings to the
spectral region of a given exciton band.

We are now in a position to explain what kind of observations made it possible
to come up with such a simple and universal expression for counting the number
of electronic excitations within a properly defined spectral region for an
arbitrary branched conjugated molecule with long enough linear segments. The
first step is bringing in the ES approach; as of today, this step is not
too surprising, since the ES approach was introduced almost a decade ago,
and is well understood by now. Still, it provides a vast simplification of the
problem, since (a) it allows one to summarize all the relevant properties of
electron interactions and correlations in a branched conjugated molecule, as
well as the details of the system Hamiltonian, by introducing relatively simple
objects, namely the excitons spectrum $\omega(k)$ and scattering matrices
$\Gamma_{a}(k)$, and (b) the problem of identification of electronic excitations in
a branched structure is reduced to solving the system of ES equations that can
be described as a generalized spectral problem in a vector space of a
relatively low dimension $n= 2n_{{\rm s}}$.

The second and most non-trivial step, is reducing the problem of solving the ES
equations to an {\it intersection problem}. This is achieved by representing
the ES equations in a form
\begin{eqnarray}
\label{ES-equations-gen} \tilde\Gamma(k) \psi = \psi.
\end{eqnarray}
where $\tilde\Gamma(k)$ is a quasimomentum dependent $2n_{{\rm s}}\times
2n_{{\rm s}}$ scattering matrix, whereas $\psi$ is a $2n_{{\rm s}}$-dimensional
vector of amplitudes of the incoming exciton waves at the scattering
centers. The matrix $\tilde\Gamma(k)$ can be represented in a natural way as a
product of three matrices [for an explicit expression see
Eq.~(\ref{ES-equations-vector})]. A reader, interested in a detailed derivation
of Eq.~(\ref{ES-equations-gen}) can find them in Section~\ref{sec:ESeqns}; here
we present some simple, intuitive arguments. Note that here we use an
abbreviated notation $\psi$ instead of $\psi^{(+)}$, used in
Section~\ref{sec:ESeqns}. The matrix $\Gamma(k)$ is block-diagonal, with the
blocks represented by $\Gamma_{a}(k)$ \footnote{Using the standard linear
algebra notation this can be represented as $\Gamma(k)=
\oplus_{a}\Gamma_{a}(k)$.}. The matrix $\Gamma(k)$ describes scattering at
molecular vertices by transforming the $2n_{{\rm s}}$-dimensional vector of
incoming exciton amplitudes to the the vector $\Gamma(k)\psi$ of the outgoing
counterparts. The diagonal matrix $\Lambda(k)$ with the eigenvalues
$\exp(ikL_{\alpha})$ describes the exciton propagation along the linear
segments. The block-diagonal matrix $P$, with $2\times 2$ blocks, accounts
for the fact that an outgoing wave, after propagation along a linear segment,
turns into an incoming wave at the opposite end of the latter. Obviously all
blocks are identical and $k$-independent. It is easy to see now that
Eq.~(\ref{ES-equations-gen}) is nothing else than a consistency condition for
a standing wave: scattering at the vertices, followed by propagation along
the segments, should result in the same wave. Stated a bit more formally,
Eq.~(\ref{ES-equations-gen}) is an alternative way of recasting the ES
equations in one of its standard forms [see Eq.~(\ref{ES-equations})].


The advantage of the representation, given by Eq.~(\ref{ES-equations-gen}), is
that the solutions of the ES equations can be considered as intersections.
Indeed, Eq.~(\ref{ES-equations-gen}) states that for a certain value $k$ of
quasimomentum, the unitary $n\times n$ (we reiterate that $n= 2n_{{s}}$) matrix
$\tilde{\Gamma}(k)$ has an eigenvalue equal to $1$. Therefore, the solutions of
the ES equations correspond to the intersections of the closed curve
$\tilde{\Gamma}(k)$, parameterized by the quasimomentum $k$, with the subspace
of $U(n)$ represented by those unitary matrices that have an eigenvalue equal
to 1, with the intersections ocurring in the space $U(n)$ of all unitary
$n\times n$ matrices.  This subspace of $U(n)$, whose matrices have an
eigenvalue equal to 1 is, hereafter referred to as $D_{1}U(n)$.  The remaining
steps would not be too surprising to someone familiar with the basics of
algebraic topology: as briefly stated in Section~\ref{sec:Intro}, we apply
topological intersection theory to compute a simpler quantity, namely the
intersection index, which provides a lower bound on the number of intersections
(i.e., the number of the solutions to the ES equation). Next, we
relate the number of ES equation solutions to the number $N$ of
electronic excitations (excitons), which involves dealing with over- and
under-counting, and finally, show the tightness of the bound when the linear
segments are long enough.

It should be noted that the standard topological intersection theory is not
directly applicable to our case, and a proper extension/generalization should
be developed. In Section~\ref{sec:top-intersection}, we present in a simple and
intuitive way the basic concepts of topological intersection theory,
rationalize why it is not directly applicable to our case, and develop a proper
extension.


\begin{figure}[tpH] \begin{center}
  \includegraphics[width=0.4\textwidth]{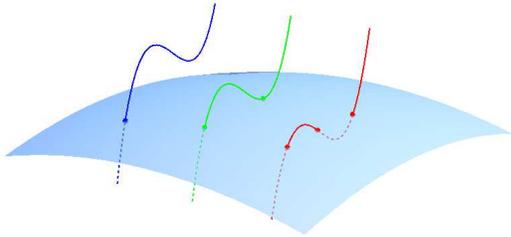} \vspace{-0.7cm}
\end{center} \caption{Solving systems of equations by means of intersection
  theory. The parametric system of equations $z = x + 8y^2 + 8y^3, x=t, 4z= -x^2 -y^2$ is solved.
  The first two equations describe a curve ($t=-1$: blue, $t=0$: green, and $t=1$: red),
 while the third equation describes a surface. We weight intersections
  with $+1$ if the curve comes from underneath or $-1$ if the curve comes from
  above, yielding an intersection index equal to $+1$.  This appreciates the
fact that we'll have at least one solution in general.} \label{decomp}
\end{figure}

\subsection{Topological Intersection Theory and its generalization}
\label{sec:top-intersection}

The main computational tool we have used in this manuscript to count the number
$N$ of electronic excitations within the spectral region of an exciton band is
topological intersection theory. In this formulation, each equation of a system
is interpreted as, generally speaking, some multi-dimensional surface, or
speaking more mathematically, a manifold. In this setting,a solution to a system of
equations manifests itself as an intersection of the aforementioned manifolds.
Finding these intersections (and specifically, the number of such) can be a
difficult task due to the generic nature of the equations. An easier
approximation to compute is a topological intersection index $q=
\sum_{j}q_{j}$, consisting of a sum over all intersections of local
intersection indices $q_{j}= \pm 1$, i.e., the intersections are weighted with
a $\pm 1$ sign factor. The topological nature of this index implies that not
only is it fixed under deformations of the curves, but also that there should
(and do) exist expressions for solutions (in our case, excitons) in purely
topological terms. Denoting the total number of intersections by $m$ (i.e.,
solutions of a system of equations), we obviously have $m \ge q$, so that
topological intersection theory provides a lower bound for the number $m$ of
solutions in terms of the topological intersection index $q$, which is much
easier to compute.

As an example, consider the family of cubics $z(t) = x(t) + 8y^2 + 8y^3$
together with $4z = -x^2 - y^2$, where $x$ is parameterized by $t$, shown in
Fig.~\ref{decomp}. We see that for different values of $t$, this family of
curves will have a different number of intersections with the surface, and
therefore the system will have a different number of solutions. However, when
intersections are weighted with $+1$ or $-1$ (depending on orientations at
crossings), we find that the sum of these weights is always 1 (the $t=0$ case
is not stable, since it bifurcates into the other cases under slight
perturbations).

At the end of Section~\ref{sec:ES-eq-main-results}, we identified the solutions
of the ES equations with the intersections of a closed curve, defined by
$\tilde{\Gamma}(k)$ in the space of the unitary group $U(n)$, with the subspace
$D_{1}(U(n))$ of unitary matrices with at least one unit eigenvalue. There is,
however, an obstacle to straightforward application of standard topological
intersection theory: the latter is valid in the case of intersecting {\it
manifolds}, namely spaces that locally look like open neighborhoods of
euclidian spaces $\mathbb{R}^{n}$. In our case, while the curve
$\tilde{\Gamma}(k)$ is a manifold, the subspace $D_{1}(U(n))$ is not, the
latter statement being due to the points, represented by unitary matrices with
a degenerate unit eigenvalues. Stated differently, the complexity of the
situation originates from degenerate solutions of the ES equations.

To treat our situation, we extend ordinary intersection theory by generalizing
a notion of the local intersection index $q_{j}$, associated with an
intersection point, so that it admits integer values, rather than just $\pm 1$.
The generalized version of $q_{j}$ is defined as follows. Consider an
intersection (equivalently a solution to the ES equation) at $k= k_{j}$, so that
$\tilde{\Gamma}(k_{j})$ has exactly $m_{j}$ unit eigenvalues; hereafter we
refer to $m_{j}$ as the multiplicity of an intersection. Recall that all
eigenvalues belong to the circle $|z|=1$ in the complex plane. In the case of
an isolated intersection (which we always have since $\tilde{\Gamma}(k)$ is an
analytic function of $k$), when $k\ne k_{j}$ is close to $k_{j}$,
$\tilde{\Gamma}(k)$ does not have unit eigenvalues. So, when $k$ goes through
$k_{j}$, starting with $k < k_{j}$ and ending up with $k > k_{j}$, there is a
certain number of eigenvalues, denoted $q_{j}$, that pass through the point $z=
1$, moving from below to above, the rest being reflected from $z=1$. We refer
to $q_{j}$ as the local intersection index; note that $q_{j}$ can be positive,
negative, or zero, and obviously $-m_{j} \le q_{j} \le m_{j}$. A more
formal definition of the local intersection index is given in
Section~\ref{sec:index-theorem}. We further define the number $m$ of the ES
equation solutions (with the degeneracy accounted for) and the global
intersection index
\begin{eqnarray} \label{define-m-and-q}
  m= \sum_{j}m_{j}, \;\;\; q= \sum_{j}q_{j},
\end{eqnarray}
so that, obviously $m \ge q$, and the intersection index provides a lower bound
for $m$.

One might ask a question: Why was the local intersection index defined as it
was? The answer is: With the above definition, the global intersection index
$q$ becomes a topological invariant, i.e., it depends only on the topological
class of $\tilde{\Gamma}(k)$. In other words, it is not changed upon
deformations of the the curve $\tilde{\Gamma}(k)$, and therefore is easily
computable. Topological invariance of $q$ is established via the Index Theorem
[Eq.~(\ref{index-theorem-intuit})].


\subsection{Winding number associated with a scattering matrix and the Index Theorem}
\label{sec:winding-index}

We start with introducing our main computational tool, namely the winding
number associated with a scattering matrix, that will allow us to derive the
Index Theorem that relates the topological intersection index, introduced in
Section~\ref{sec:top-intersection}, to the winding number of the scattering
matrix $\tilde{\Gamma}(k)$, as well as explicitly compute the latter. Hereafter
we use the term scattering matrix for any quasimomentum $k$ dependent $n\times n$
unitary matrix $f(k)$, where $n$ can be arbitrary. Examples of scattering
matrices are $r_{a}\times r_{a}$ scattering matrices $\Gamma_{a}(k)$ at the
molecular vertices, as well as $2n_{{\rm s}}\times 2n_{{\rm s}}$ matrices
$\tilde{\Gamma}(k)$, $\Gamma(k)$, $\Lambda(k)$, and $P$, introduced in
Section~\ref{sec:ES-eq-main-results}, as well as, in a more formal way in
Section~\ref{sec:ESeqns}. Note that $P$ is actually $k$-independent. The
integer-valued winding number $w(f)$ is defined by making use of the fact that
the determinant of a unitary matrix is a unimodular complex number, i.e., we
can represent $\det f(k)= e^{i\varphi(k)}$:
\begin{eqnarray}
\label{define-w-intuitive} w(f) &=& \int_{-\pi}^{\pi}\frac{dk}{2\pi i}(\det f(k))^{-1}\frac{d}{dk}\det f(k) \nonumber \\ &=& \int_{-\pi}^{\pi}\frac{dk}{2\pi i}\frac{d\ln\left(\det f(k)\right)}{dk}= \int_{-\pi}^{\pi}\frac{dk}{2\pi }\frac{d\varphi(k)}{dk}.
\end{eqnarray}
Since $\varphi$ is a phase (angular) variable, i.e., it is defined up to an
integer multiple of $2\pi$, the integral in the r.h.s. of
Eq.~(\ref{define-w-intuitive}) is an non-zero integer number; stated
differently, it reflects the multi-valued nature of the logarithm, as a function
of a complex argument.

Explicit computation of the generalized intersection index $q$, introduced in
Section~\ref{sec:top-intersection}, is possible due to the following algebraic
property of the winding number. Consider scattering matrices $f(k)$, $g(k)$,
and $h(k)$ of the sizes $n\times n$, $n\times n$, and $m\times m$,
respectively. Denote by $f\cdot g$ and $f\oplus h$ the  $n\times n$ matrix
product of $f$ and $g$, and the $(n+m)\times (n+m)$ block-diagonal matrix with
the blocks, given by $f$ and $h$, respectively. The well-known multiplicative
properties of the determinantm, namely $\det (f\cdot g)= (\det f)(\det g)$ and
$\det(f\oplus h)= (\det f)(\det h)$, immediately imply the desired algebraic
properties of the winding number
\begin{eqnarray}
\label{w-algebraic-property} w(f\cdot g)= w(f)+ w(g), \;\; w(f\oplus h)= w(f)+ w(h).
\end{eqnarray}

The algebraic properties [Eq.~(\ref{w-algebraic-property})] allow for a concise
computation of the winding number $w(\tilde{\Gamma})$.
\begin{eqnarray}
\label{compute-w-tilde-G} w(\tilde{\Gamma})= w(\Lambda)+ w(\Gamma)+ w(P)= 2\sum_{\alpha}L_{\alpha}+ \sum_{a}w(\Gamma_{a})
\end{eqnarray}
In deriving Eq.~(\ref{compute-w-tilde-G}), we have used the fact that $P$ is
$k$-independent, which due to Eq.~(\ref{define-w-intuitive}), implies $w(P)=0$.
A direct calculation shows $w(\Lambda_{\alpha})=L_{\alpha}$
for an $1\times 1$ scattering matrix $\Lambda_{\alpha}(k)= e^{ikL_{\alpha}}$.
Recalling the definition $l_{a}= w(\Gamma_{a})$, we arrive at
\begin{eqnarray}
\label{w-tilde-G-final} w(\tilde{\Gamma})=  2\sum_{\alpha}L_{\alpha}+ \sum_{a}l_{a}, \;\;\; l_{a}\equiv w(\Gamma_{a}).
\end{eqnarray}

The lower bound
\begin{eqnarray}
\label{bound-explicit-intuit} m =\sum_{j}m_{j} \ge \sum_{j}q_{j}=q = 2\sum_{\alpha}L_{\alpha} + \sum _{a}l_{a}
\end{eqnarray}
for the number of solutions to the ES equations, which becomes tight for long enough
segments, is obtained by combining the explicit expression given by
Eq.~(\ref{w-tilde-G-final}) with the Index Theorem
\begin{eqnarray}
\label{index-theorem-intuit} q= \sum_{j}q_{j}=w(\tilde{\Gamma}).
\end{eqnarray}
Note that the tightness of the bound [Eq.~(\ref{bound-explicit-intuit})]
follows from the statement that $q_{j}= m_{j}$ for long enough
segments. We comment on the above statement in
Section~\ref{sec:ES-solution-to-exciton}, and provide a more formal derivation
in Section~\ref{sec:Long-arm}.

We conclude this subsection with simple intuitive arguments in support of the Index Theorem [Eq.~(\ref{index-theorem-intuit})], a sketch of a formal proof, for a reader, interested in details, is presented in Section~\ref{sec:sketch-proof-index}.

The intuition relies on two equivalent representations of charged particles
current/flux. Consider the eigenvalues $\lambda_{1},\ldots,\lambda_{n}$ (with
possible degeneracy properly accounted for) of $\tilde{\Gamma}$ as a system of
$n$ undistinguishable particles with charge $+1$ that reside in a unit circle
$|z|=1$ in the complex plane $\mathbb{C}$, and let us think of $k$ as time.
Then the scattering matrix $\tilde{\Gamma}(k)$ provides a periodic trajectory
of our $n$-particle system. To calculate the total charge flux in the system
over a period, we count the total number of full rotations of all particles in
the system. Since $\det{\tilde{\Gamma}(k)}=\prod_{i=1}^{n}\lambda_{i}(k)$ with
$\lambda_{i}(k)$ being the eigenvalues of $\tilde{\Gamma}(k)$, the total flux,
due to Eq.~(\ref{w-algebraic-property}) is given by $q= \sum_{j}q_{j}=
\sum_{j}w(\lambda_{j})= w(\tilde{\Gamma})$. On the other hand, the total charge
flux $q$ is given by the total charge that went through any cross section,
e.g., the total charge traveled through the point $z=1$ of the circle. For the
second interpretation, the charge travels through $z= 1$ at times $k=k_{j}$,
and, according to the definition of the local intersection index, given in
Section~\ref{sec:top-intersection}, in the amount of $q_{j}$, so that
$q=\sum_{j}q_{j}$. The index theorem is a reflection of the fact that these two
interpretations of charge flux are the same (taking orientation into account).

\begin{figure}[tph]
  \begin{center}
    \includegraphics[width=0.5\textwidth]{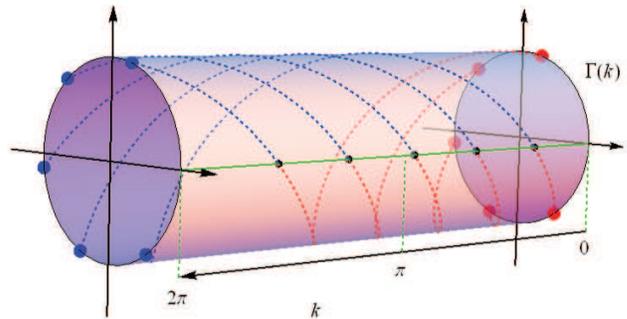}
  \end{center}
  \vspace{-0.6cm}
  \caption{Two equivalent interpretations of the flux that result in the index theorem. The eigenvalues
    form
    a periodic trajectory, evolving over time ($k$),
    where each eigenvalue has charge $+1$. To calculate the total
    flux, one can count the full number of rotations of all particles, which
    amounts to calculating $w(\tilde \Gamma)$. Alternatively, one could
    calculate the total charge passing through any cross-section, e.g. $z= 1$ (green line), yielding $q= \sum q_j$. The comparison of the two
    interpretations gives the index theorem [Eq.~(\ref{index-theorem-intuit})].}
 \label{fig:index}
\end{figure}

\subsection{From solutions of the  ES equation to excitons and the long-arm limit}
\label{sec:ES-solution-to-exciton}

In this section, we finalize our computation by relating the number of
solutions of the ES equations $m$ to the actual number of electronic
excitations $N$ within the spectral region defined by the exciton band, which
results in the lower bound for $N$ [Eq.~(\ref{exciton-bound-general})].
We further comment on the tightness of the bound for the case of long enough
segments.

We start by noting that any exciton corresponds to a solution of
the ES equation. On the other hand, if we have a solution of the ES equation
with $k= k_{j}$, we should expect to have a solution at $k= -k_{j}$,
corresponding to the same exciton. This is evident since the latter is represented by a standing
wave, so that we have two plane waves at each linear segment, propagating in
the opposite directions. More precisely, these two plane waves have opposite
quasimomenta, due to time-reversal symmetry. The above statement is formalized
in Section~\ref{sec:ESeqns}, where we demonstrate that if $\psi$ is a solution
of the ES equation [Eq.~(\ref{ES-equations-gen})] for $k= k_{j}$, then
$\Gamma(k_{j})\psi$ is a solution with $k= -k_{j}$, which corresponds to the
same exciton. This implies that the solutions of the ES equation with $k\ne 0,\pi$
count each exciton twice. The situation when $k= 0,\pi$ is a just a bit more
complicated. In particular, due to $k= -k$ for $k=0, \pi$, the two plane waves on each
segment are actually one wave. As demonstrated in Section~\ref{sec:ESeqns}, for
$k= 0,\pi$, the vector space of solutions to the ES equation is invariant with
respect to the action of $\Gamma(k)$, and $(\Gamma(k))^{2}=1$. This implies
that the bandedge, i.e., $k= 0,\pi$, solutions have a well-defined parity,
i.e., even and odd solutions with $\Gamma(k)\psi= \psi$, and $\Gamma(k)\psi=
-\psi$. Denote the number of even and odd solutions by $d_{k}^{+}$ and
$d_{k}^{-}$, respectively, and further introduce the bandedge indices $d_{k}=
d_{k}^{+}- d_{k}^{-}$. It is easy to understand that the odd solutions are
unphysical, since they correspond to a complete zero standing wave (the plane
waves with $\pm k$ cancel each other), whereas even solutions correspond to
actual excitons. Therefore, we arrive at a simple relation
\begin{eqnarray}
\label{m-to-N-intuitive}
N=\frac{m+ d_{0}+ d_{\pi}}{2}
\end{eqnarray}
between the number of electronic excitations $N$ within the spectral region of
an exciton band and the number of solutions to the ES equations $m$. The main
counting result [Eq.~(\ref{exciton-bound-general})] is readily obtained by
simply combining Eqs.~(\ref{bound-explicit-intuit}) and
(\ref{m-to-N-intuitive}).

The tightness of the bound for long enough linear segments is due to the fact
that in this limit, $q_{j}= m_{j}$. According to the definition of the
local intersection index, this means that if we have an $m_{j}$-fold degenerate
solution of the ES equations for $k= k_{j}$, i.e., $\tilde{\Gamma}(k_{j})$ has
exactly $m_{j}$ eigenvalues equal to $1$, they all move above and below the
point $z=1$ on their circle of residence, when $k$ becomes slightly larger and
slightly smaller than $k_{j}$, respectively. This is demonstrated in
Section~\ref{sec:Long-arm} by applying standard quantum mechanical
perturbation theory.

\section{Details of the Derivation}
\label{sec:Derivation-details}

In this section we present the details of the derivation of our results on counting electronic excitations in branched conjugated molecules.

\subsection{The ES equations}
\label{sec:ESeqns}

We start by considering a quasi-one-dimensional quantum system on a graph (a
collection of vertices connected by edges), whose states (excitons) are
quasiparticles \cite{SGD:polymer, ScholesGD:Excns, ES-naturephys}. The latter
are represented by plane waves that reside on the linear segments (graph edges
$\alpha$), getting scattered at the scattering centers (graph vertices $a$),
which leads to the formation of standing waves that describe discrete quantum
states of the system. We consider the case when the linear segments possess
discrete translational symmetry, which is broken by the finite segment length
only. Therefore, an exciton in an infinite chain has a well-defined quantum
number, namely quasimomentum $k\in [-\pi,\pi]$ that resides in the Brillouin
zone, and the exciton properties are described by its spectrum $\omega(k)$
\cite{ES-prl,ES1,ES2}. The time-reversal symmetry implies
$\omega(-k)=\omega(k)$. We also assume that for all $\omega_{0}$ inside the
exciton band (excluding the band edges $k=0,\pi$) there are exactly two values
of quasimomentum, i.e., $\pm k_{0}$ so that $\omega(\pm k_{0})=\omega_{0}$.

Every scattering center $a$ is described by a frequency-dependent scattering
matrix $\Gamma_{a}$ that relates the amplitudes of the outgoing plane waves to
the incoming counterparts \cite{ES-prl,ES1}. The scattering matrices
$\Gamma_{a}(k)$ are unitary ($\Gamma_{a}(-k)=\Gamma_{a}^{\dagger}(k)$) and
$\Gamma_{a}(k)\Gamma_{a}(-k)=1$, since the quasimomenta of the incoming and
outgoing waves are different just by the sign due to time-reversal symmetry.
The scattering matrix can be viewed as a map $\Gamma_{a}:S^{1}\to U(n_{a})$ of
the circle $S^{1}\subset \mathbb{C}$, naturally embedded into the complex plane
(by $z=e^{ik}$), that represents the Brillouin zone to the unitary group, with
$n_{a}$ being the vertex degree (the number of edges attached to it). The
scattering matrices are analytical functions of $z\in S^{1}$, which implies
they are complex analytical functions of $z$ in some neighborhood of $S^{1}$ in
$\mathbb{C}$.

Denoting by $\psi_{a\alpha}^{(\pm)}$ the amplitudes of the incoming/outgoing
waves at vertex $a$ from/to edge $\alpha$, the ES equations are given by
\begin{eqnarray} \label{ES-equations}
  \psi_{a\alpha}^{(-)}=\sum_{\beta}\Gamma_{a,\alpha\beta}(k)\psi_{a\beta}^{(+)},
  \;\;\; \psi_{a\alpha}^{(+)}=e^{ikL_{\alpha}}\psi_{b\alpha}^{(-)},
\end{eqnarray} where $L_{\alpha}\in \mathbb{N}$ is the (integer) length of a
segment $\alpha$, and in the second set of equations $\{a,b\}$ is the border of
edge $\alpha$. The first set of equations in~(\ref{ES-equations}) connects the
amplitudes of the outgoing waves to the incoming counterparts at scattering
centers, whereas the second set describes the change of the wave phase as a
result of traveling along a linear segment.

We can further treat
$\psi^{(\pm)}=\sum_{a\alpha}\psi_{a\alpha}^{(\pm)}e_{a\alpha}$ as vectors in
the vector space $V$, spanned on the set of oriented vertices $e_{a\alpha}$ of
our graph. We naturally represent
$V=\oplus_{a}V_{a}=\oplus_{\alpha}V_{\alpha}$, where $V_{a}$ and $V_{\alpha}$
are the vector spaces spanned on the vectors $\{e_{a\alpha}\}$ ($\alpha$ is
attached to $a$) and $\{e_{b\alpha},e_{c\alpha}\}$ ($\alpha$ connects $b$ and
$c$), respectively. Thus $\Gamma_{a}(k)$ is a family of unitary matrices
parameterized by $k$ acting in $V_{a}$. This allows us to define a family
$\Gamma(k)=\oplus_{a}\Gamma_{a}(k)$ of unitary matrices acting in the space
$V$. We further introduce a unitary matrix $P=\oplus_{\alpha}P_{\alpha}$ acting
in $V$ with $P_{\alpha}e_{b\alpha}=e_{c\alpha}$ when $\alpha$ connects $b$ and
$c$, and a family of unitary matrices
$\Lambda(k)=\oplus_{\alpha}e^{ikL_{\alpha}}{\rm id}_{V_{\alpha}}$. Using the
above notations, the ES equations can be combined into one vector equation for
$\psi^{(+)}$ only \begin{eqnarray} \label{ES-equations-vector}
  \tilde{\Gamma}(k)\psi^{(+)}=\psi^{(+)}, \;\;\; \tilde{\Gamma}(k)\equiv
  \Lambda(k)P\Gamma(k), \end{eqnarray} and we can interpret
  $\tilde{\Gamma}:S^{1}\to U(n)$, with $n=\sum_{a}n_{\alpha}$, as a map of the
  Brillouin zone to the unitary group $U(n)$ (see the appendix for a brief
  overview of topological properties of the unitary group). Representing the ES
  equations in the form~(\ref{ES-equations-vector}) is crucial in obtaining the
  main result of this manuscript.

Since $\tilde{\Gamma}(k)$ is an analytic function of quasimomentum $k$,
Eq.~(\ref{ES-equations-vector}) can hold only for a finite number of
quasimomenta $k_{j}\in S^{1}$ (hereafter $k$ is identified with $z=e^{ik}$). We
further denote the number of linearly independent solutions of
Eq.~(\ref{ES-equations-vector}) for $k=k_{j}$ by $m_{j}$, referred to as the
multiplicity. It is natural to refer to $m=\sum_{j}m_{j}$ as the total number
of solutions of the ES equations (accounting for the degeneracy). Our strategy
is (i) finding a relation between $m$ and the number of quantum
states/excitons, and (ii) establishing a lower bound for $m$, as well as
showing that the bound is tight when the segment lengths $L_{\alpha}$ are large
enough. Task (i) involves elementary linear algebra only. Achieving task (ii)
necessitates topological intersection theory and constitutes the main technical
result of this manuscript.

If $\psi^{(+)}$ is a solution of Eq.~(\ref{ES-equations-vector}) with $k= k_{j}$, so that $k_{j} \ne
0,\pi$, then $\Gamma(k_{j})\psi^{(+)}$ is also a solution with $k= -k_{j}$, which follows from
$\tilde{\Gamma}(-k_{j})\Gamma(k_{j})\psi^{(+)}=
\Lambda(-k_{j})P\psi^{(+)}=P\Lambda(-k_{j})\psi^{(+)}=
\Gamma(k_{j})\psi^{(+)}$. Here we have used $\Gamma(-k)\Gamma(k)=1$,
$\Lambda(-k)\Lambda(k)=1$, $P^{2}=1$, and $[P,\Lambda(k)]=0$. Obviously these
two solutions represent the same standing wave, i.e., the same exciton.
Therefore the number of excitons with $k\ne 0,\pi$ is equal to the half of
number of solutions of Eq.~(\ref{ES-equations-vector}) with $k\ne 0,\pi$. The
cases of $k=0,\pi$ need a bit more care. First of all, a definition of
$\Gamma_{a}(0)$ and $\Gamma_{a}(\pi)$ is not obvious, since in these cases we
have $k=-k$, and there is no distinction between the incoming and outgoing
waves. We can still define $\Gamma_{a}(0)$ and $\Gamma_{a}(\pi)$ by using the
analytical continuation from the Brillouin zone with $k\ne 0,\pi$. This leads
to the matrices $\Gamma(k)\in U(n)$ with $k=0,\pi$ that satisfy
$(\Gamma(k))^{2}=1$. Introducing $\tilde{P}(k)=\Lambda(k)P$, we have
$(\tilde{P}(k))^{2}=1$. Therefore Eq.~(\ref{ES-equations-vector}) in the cases
$k=0,\pi$ can be recast in an equivalent form
$\Gamma(k)\psi^{(+)}=\tilde{P}(k)\psi^{(+)}$.

The excitons \cite{SGD:polymer, ScholesGD:Excns, ES-naturephys} (i.e., quantum
states, rather than the solutions of the ES equations) are described by a vector $\psi\in V$ of amplitudes, rather than a pair $\psi^{(\pm)}$ of vectors (due to $k=-k$), and the equations adopt a form $\Gamma(k)\psi=\psi$, $\tilde{P}(k)\psi=\psi$, which can be recast as
\begin{eqnarray}
\label{ES-equations-bandedge} \Gamma(k)\psi=\tilde{P}(k)\psi, \;\;\; \Gamma(k)\psi=\psi.
\end{eqnarray}
The first equation~(\ref{ES-equations-bandedge}) is the $k=0,\pi$ version of the ES equations. Therefore, the vector spaces $W_{k}^{+}\subset W_{k}$ of the bandedge excitons are the subspaces of the spaces $W_{k}$ of solutions to the ES equation that satisfy the second equation~(\ref{ES-equations-bandedge}). It is straightforward to verify that $W_{k}$ is an invariant subspace of $\Gamma(k)$. Therefore we have $W_{k}=W_{k}^{+}\oplus W_{k}^{-}$, where $W_{k}^{-}$ is the unphysical space defined by the condition $\Gamma(k)\psi=-\psi$. These correspond to solutions that provide zero everywhere wavefunctions of the standing waves (excitons). Introducing $d_{k}^{\pm}=\dim W_{k}^{\pm}$ and recalling our consideration of the $k\ne 0,\pi$ case, we obtain for the number of excitons $N$
\begin{eqnarray}
\label{number-of-excitons}
N=\frac{1}{2}\left(m+(d_{0}^{+}-d_{0}^{-})+(d_{\pi}^{+}-d_{\pi}^{-})\right).
\end{eqnarray}
Two comments are in place. First, identification of $d_{k}^{\pm}$ is much easier compared to finding $m$, since it is a linear problem, compared to a generalized spectral problem. Second, in a generic situation we have $\Gamma_{a}(k)=-1$ for $k=0,\pi$, which yields $d_{k}^{+}=0$, $d_{k}^{-}=n/2= n_{{\rm s}}$.

\subsection{The Index Theorem}
\label{sec:index-theorem}

To establish a topological bound for the multiplicity $m$, we need to introduce
a natural topological invariant, hereafter referred to as the winding number.
Any continuous function $S^1 \rightarrow S^1$ has a natural topological
invariant, known as its winding number. Intuitively, it counts how many times
the domain $S^1$ wraps around the codomain $S^1$. Due to continuity, this must
be an integer number (taking orientation into account). In our case, this
integer $w(f)$, associated with a map $f:S^{1}\to U(n)$ of the Brillouin zone
to the unitary group, is defined by
\begin{eqnarray}
\label{define-w} w(f)=\int_{-\pi}^{\pi}\frac{dk}{2\pi i}(\det f(k))^{-1}\frac{d}{dk}\det f(k).
\end{eqnarray}
An interpretation of $w(f)$ that demonstrates its integer
character is as follows. For a described map $f$, we have a map $\det
f:S^{1}\to U(1)$, since the determinant of a unitary matrix is a unimodular
number. Then $w(f)$, defined by Eq.~(\ref{define-w}), is the winding number of
$\det f:S^1 \rightarrow U(1)$, i.e., the number of times $\det f(k)$ winds over
the circle $U(1)\cong S^{1}$, while $k$ goes once over the Brillouin zone
$S^{1}$. The winding number is a topological (homotopy) invariant, meaning it
does not change upon continuous deformations of the map $f$. In our further derivations we will make substantial use the following key algebraic properties of the winding number, outlined in Section~\ref{sec:winding-index}: For any maps $f,g:S^{1}\to
U(n)$ and $h:S^{1}\to U(m)$, the relations, given by Eq.~(\ref{w-algebraic-property}) hold. We reiterate that
$f\cdot g:S^{1}\to U(n)$ and $f\oplus h:S^{1}\to U(n+m)$ are obtained by point-wise multiplication of matrices and forming a block-diagonal matrix, respectively, as well as that the relations of Eq.~(\ref{w-algebraic-property}) are a direct consequence of the properties of determinants, namely $\det(f\cdot g)=(\det f)(\det g)$ and $\det(f\oplus h)=(\det f)(\det h)$, combined with the definitions of Eq.~(\ref{define-w}).

To formulate the topological bound, we begin with a naive observation of any
solution $k=k_j$ of the ES equations [Eq.~(\ref{ES-equations-vector})]:
$\tilde{\Gamma}(k_{j})$ should be unitary and have an eigenvalue equal to $1$.
If we let $D_1U(n)$ denote the set of $n \times n$ unitary matrices that have
at least one eigenvalue equal to 1, then solutions of the ES equations are
associated with the intersections of the $1$-dimensional cycle in $U(n)$
defined by the map $\tilde{\Gamma}:S^{1}\to U(n)$ with the
$(n^{2}-1)$-dimensional subspace $D_{1}U(n)\subset U(n)$.  Note that $\dim
U(n)=n^{2}$.  To compute $m$, the number of solutions of the ES equation, we need to
count the intersections weighted with their multiplicities $m_{j}$. To do so,
we further introduce the intersection index, that counts the intersections
weighted with local intersection indices $q_{j}$. Generally, $q_j$ is different
from $m_{j}$, whereas the total index $q=\sum_{j}q_{j}$ is a topological
invariant, meaning it depends only on the topological classes of the
intersecting cycles, in our case $D_{1}U(n)$ and $\tilde{\Gamma}$. A statement
that explicitly computes $q$ as a topological invariant and relates it to the
sum of the local intersection indices is usually referred to as an {\it index
theorem}~\cite{Spanier:AT}. In our case, it adopts the form 
\begin{eqnarray}
  \label{index-theorem} \sum_{j}q_{j}=w(\tilde{\Gamma}).  
\end{eqnarray}
In what follows we (a) define the local indices $q_{j}$ and present plausible
arguments on the validity of Eq.~(\ref{index-theorem}), (b) show that
$|q_{j}|\le m_{j}$, (c) further show that when the segment lengths $L_{\alpha}$
are large enough we have $q_{j}=m_{j}$, and (d) explicitly compute 
  \begin{eqnarray} \label{compute-w(Gamma)}
    w(\tilde{\Gamma})=2\sum_{\alpha}L_{\alpha} + \sum _{a}l_{a}, \;\;\;
    l_{a}\equiv w(\Gamma_{a}).  
  \end{eqnarray} 
Combining (b), (c), and Eq.~(\ref{compute-w(Gamma)}) with
Eq.~(\ref{index-theorem}) we obtain an explicit bound 
\begin{eqnarray}
  \label{bound-explicit} m\ge 2\sum_{\alpha}L_{\alpha} + \sum _{a}l_{a}
\end{eqnarray} 
that becomes tight for long enough segments.

Derivation of Eq.~(\ref{compute-w(Gamma)}) is straightforward. We start with a
representation for $\tilde{\Gamma}$ given by Eq.~(\ref{ES-equations-vector})
with explicit expressions for the three factors in the r.h.s.  \begin{eqnarray}
  \label{tilde-gamma-explicit}
  \tilde{\Gamma}=\left(\oplus_{\alpha}e^{ikL_{\alpha}}{\rm
  id}_{V_{\alpha}}\right)\cdot \left(\oplus_{\alpha}P_{\alpha}\right)\cdot
  \left(\oplus_{a}\Gamma_{a}\right) \end{eqnarray} and apply the properties of
  the winding number [Eq.~(\ref{w-algebraic-property})], combined with explicit
  computations $w\left(e^{ikL_{\alpha}}{\rm
  id}_{V_{\alpha}}\right)=2L_{\alpha}$, $w(P_{\alpha})=0$ and the definition of
  $l_{a}$, given in Eq.~(\ref{compute-w(Gamma)}).

The local intersection indices are defined in the following way. Consider a a
solution of Eq.~(\ref{ES-equations-vector}) [or equivalently an intersection of
  $\tilde \Gamma$ with $D_1U(n)$] of multiplicity $m_{j}$ at $k=k_{j}$. The
  analytic nature of $\tilde \Gamma(k)$ implies solutions are isolated.
  Therefore, there is a small neighborhood
  $(k_{j}-\varepsilon,k_{j}+\varepsilon)$ of $k_{j}$ for which there are no
  other solutions.  For any $k$ in this neighborhood, there will be $m_{j}$
  eigenvalues of $\tilde{\Gamma}(k)$ that are close to $1$ that can be
  distinguished from the other eigenvalues. If $k \neq k_j$, then all of these
  $m_j$ eigenvalues must be close, but not equal, to 1 (any eigenvalue
  precisely equal to 1 would signify a solution and contradict the isolated
  nature of the solutions).  Let $m_{j}^{+}$ be the number of eigenvalues close
  to 1 with positive imaginary part for $k \in (k_j , k_j+ \varepsilon)$ and
  $m_{j}^{-}$ be the number of eigenvalues with positive imaginary part for
  $k\in (k_{j} - \varepsilon,k_{j})$.  Define the local intersection index $q_j
  := m_j^+ - m_j^-$. Naturally, $q_{j}$ can be interpreted as the number of
  eigenvalues of $\tilde{\Gamma}(k)$ that move through the point $1\in U(1)$ in
  the counterclockwise direction while $k \in (k_j-\varepsilon, k_j +
  \varepsilon)$ goes through $k_{j}$ in the counterclockwise direction.
  Obviously, $|q_{j}|\le m_{j}$.

\subsection{Sketch of a proof of Index Theorem}
\label{sec:sketch-proof-index}

Intuitive arguments in support of validity of the Index Theorem [Eq.~(\ref{index-theorem})] have been presented at the end of Section~\ref{sec:winding-index}.
In this section we present a sketch of a proof. Consider a map $f:S^{1}\to U(n)$ that intersects $D_{1}U(n)$ at a finite number of (isolated) points $k_{j}\in S^{1}$. The main idea is to replace the segments of the curve $f:S^{1}\to U(n)$ for the values of the parameter $k_{j}-\varepsilon< k< k_{j}+\varepsilon$ close to the intersection points $k_{j}$ with the segments that miss $D_{1}U(n)$ and compute the change in the winding number due to the aforementioned replacement. The resulting curve
$\tilde{f}$ misses $D_{1}U(n)$, so that $w(\tilde{f})=0$, and the winding number $w(f)$ is described by the total change of the winding number due to the all replacements.

Note that $f(k_{j})$ has exactly $m_{j}$ unit eigenvalues. Denote
$Y_{j}^{\varepsilon}=\{k\in S^{1}: k_{j}-\varepsilon< k<
k_{j}+\varepsilon\}$, and $Z_{\delta}=\{e^{i\varphi}:|\varphi|< \delta\}$.
Obviously we can choose $\varepsilon> 0$ and $\delta> 0$ to be small enough,
so that $f(k)$ has exactly $m_{j}$ eigenvalues that belong to $Z_{\delta}$
for all $k\in Y_{j}^{\varepsilon}$, and $Y_{j}^{\varepsilon}$ do not
intersect for different $j$. Choose some $k_{j}^{-},k_{j}^{+}\in
Y_{j}^{\varepsilon}$ so that $k_{j}^{-}< k_{j}< k_{j}^{+}$ and replace the
curve segments $f|_{[k_{j}^{-},k_{j}^{+}]}: [k_{j}^{-},k_{j}^{+}]\to U(n)$
with the curves $\tilde{f}_{j}: [k_{j}^{-},k_{j}^{+}]\to U(n)$ that miss
$D_{1}U(n)$ in the following way. Fix basis sets $e_{i}^{(j)\pm}$ that
diagonalizes the matrices $f(k_{j}^{\pm})$ and let $\lambda_{i}^{(j)\pm}$ be
the corresponding eigenvalues. Consider two sets of the eigenvalue
trajectories $\lambda_{i}^{(j)}:[k_{j}^{-},k_{j}^{+}]\to U(1)$ that start
and end at $\lambda_{i}^{(j)-}$ and $\lambda_{i}^{(j)+}$, respectively. Note
that any set of the eigenvalue trajectories can be extended to a path
$[k_{j}^{-},k_{j}^{+}]\to U(n)$ that starts and ends at $f(k_{j}^{-})$ and
$f(k_{j}^{+})$, respectively by choosing basis set trajectories $e^{(j)}:
[k_{j}^{-},k_{j}^{+}]\to U(n)$ (here we identify the sets of orthonormal
basis sets with the elements of the unitary group $U(n)$) that start and end
at $e^{(j)-}$ and $e^{(j)+}$, respectively. Denote these two extensions by
$f_{j}^{(r)}:[k_{j}^{-},k_{j}^{+}]\to U(n)$ with $r=0,1$. The first set of
the eigenvalue trajectories connect the initial eigenvalues to their final
counterparts via the paths of the minimal length. The corresponding
extensions $f_{j}^{(0)}$ to paths in $U(n)$ are topologically (homotopy)
equivalent to the original path segments $f|_{[k_{j}^{-},k_{j}^{+}]}$. The
second set connects the initial to the final eigenvalues in a way that the
trajectories miss the point $\lambda=1$. This can be achieved in the
following way. Among the first set of trajectories there are exactly
$|q_{j}|$ ones that go through the point $\lambda=1$ to change the number of
eigenvalues, which lie in $Z_{\delta}$ and have a positive imaginary part,
from $m_{j}^{-}$ to $m_{j}^{+}$. We replace these $|q_{j}|$ trajectories
with the ones that go over the circle in the opposite direction, missing the point
$\lambda=1$.

The contributions of both types of trajectory segments to the integral
representation for the winding number [Eq.~(5)] are
\begin{eqnarray}
\label{w-segments} w(f_{j}^{(r)})= \int_{k_{j}^{-}}^{k_{j}^{+}}\frac{dk}{2\pi}(\det{f_{j}^{(r)}(k)})^{-1}\frac{d}{dk}\det{f_{j}^{(r)}(k)} \nonumber \\ =\sum_{i}\int_{k_{j}^{-}}^{k_{j}^{+}}\frac{dk}{2\pi}(\lambda_{i}^{(j,r)}(k))^{-1}
\frac{d}{dk}(\lambda_{i}^{(j,r)}(k)).
\end{eqnarray}
Comparing the integrals under the summation sign in Eq.~(\ref{w-segments})
for $r=1$ and $r=0$, we can see that they are identical for all $i$, except
for those that correspond to the eigenvalue trajectories that were replaced
to avoid going through the $\lambda=1$ point. In the latter case, the
difference between two integrals between the $r=1$ and $r=0$ cases can be
represented as an integral over the whole circle with the eigenvalue winding
exactly once over $U(1)$ in the clockwise (counterclockwise) direction for
the replaced segment that corresponds to going through the $\lambda=1$ point
in the counterclockwise (clockwise) direction, respectively. This yields
\begin{eqnarray}
\label{w-local-difference} w(f_{j}^{(1)})-w(f_{j}^{(0)})=-q_{j}.
\end{eqnarray}

Denote by $\tilde{f}:S^{1}\to U(n)$ the curve obtained by replacing the
original segments $f|_{[k_{j}^{-},k_{j}^{+}]}$ with $f_{j}^{(1)}$. As shown
earlier, replacing the original segments with $f_{j}^{(0)}$ does not change
the winding number. Therefore it follows from Eq.~(\ref{w-local-difference})
that
\begin{eqnarray}
\label{w-global-difference} w(\tilde{f})-w(f)=-\sum_{j}q_{j}.
\end{eqnarray}
Since $\tilde{f}$, by construction, misses $D_{1}U(n)$, we have
$w(\tilde{f})=0$, so that Eq.~(\ref{w-global-difference}) implies the
statement of the index theorem [Eq.~(\ref{index-theorem})].

A rigorous proof of the index theorem can be given using homological arguments, which
is beyond the scope of this manuscript. This formal argument will be given in a later paper.

\subsection{The long arm limit}
\label{sec:Long-arm}

To demonstrate the tightness of the bound in Eq.~(\ref{bound-explicit}) we
show that for long enough segments at any intersection, we have
$m_{j}^{+}=m_{j}$ and $m_{j}^{-}=0$, so that $q_{j}=m_{j}$, which validates
(c). This can be done by applying quantum mechanical perturbation theory in the
degenerate case, i.e., in a small neighborhood of an intersection $k_{j}$
\begin{eqnarray}
\label{perturb-Gamma} \tilde{\Gamma}(k)\approx \tilde{\Gamma}(k_{j})&+&(k-k_{j})\frac{d\Lambda(k_{j})}{dk_{j}}\Lambda^{-1}(k_{j})\tilde{\Gamma}(k_{j}) \nonumber \\ &+& (k-k_{j})\tilde{\Gamma}(k_{j})\Gamma^{-1}(k_{j})\frac{d\Gamma(k_{j})}{dk_{j}}.
\end{eqnarray}
The first term in the perturbation grows linearly with the segment length,
whereas the second one is segment length independent and can be neglected
for long enough segments. So that Eq.~(\ref{perturb-Gamma}) can be approximated as
\begin{eqnarray}
\label{perturb-Gamma-2} d\tilde{\Gamma}(k)\approx i(k-k_{j})L\tilde{\Gamma}(k_{j}),
\end{eqnarray}
where $L=\oplus_{\alpha}L_{\alpha}{\rm id}_{V_{\alpha}}$ is a
Hermitian operator whose eigenvalues are the segment lengths
$L_{\alpha}$. According to quantum mechanical perturbation theory,
the first-order correction to a (possibly degenerate) eigenvalue is
given by the eigenvalues of the projection of the perturbation to
the operator onto the subspace of eigenvectors that correspond to
the zero-order eigenvalue, which in our case is the
$m_{j}$-degenerate unit eigenvalue. Since $\tilde{\Gamma}(k_{j})$
acts as the unit operator in the relevant eigenvector subspace we
need to inspect the eigenvalues of the projection of $L$. According
to the quantum mechanical variational principle, all eigenvalues of
the projection of a Hermitian operator exceed the lowest eigenvalue
of the operator itself, which in the case of $L$ is the minimal
segment length. Therefore, in a small neighborhood of $k_{j}$, where
the perturbation theory is applicable, the imaginary part of all
close to one eigenvalues is positive/negative for $k>k_{j}$ and
$k<k_{j}$, respectively, which completes the argument.

\section{Application to Polyflourene-based molecule}
\label{sec:Application}

\begin{figure}[tph]
  \begin{center}
    \includegraphics[width=0.25\textwidth]{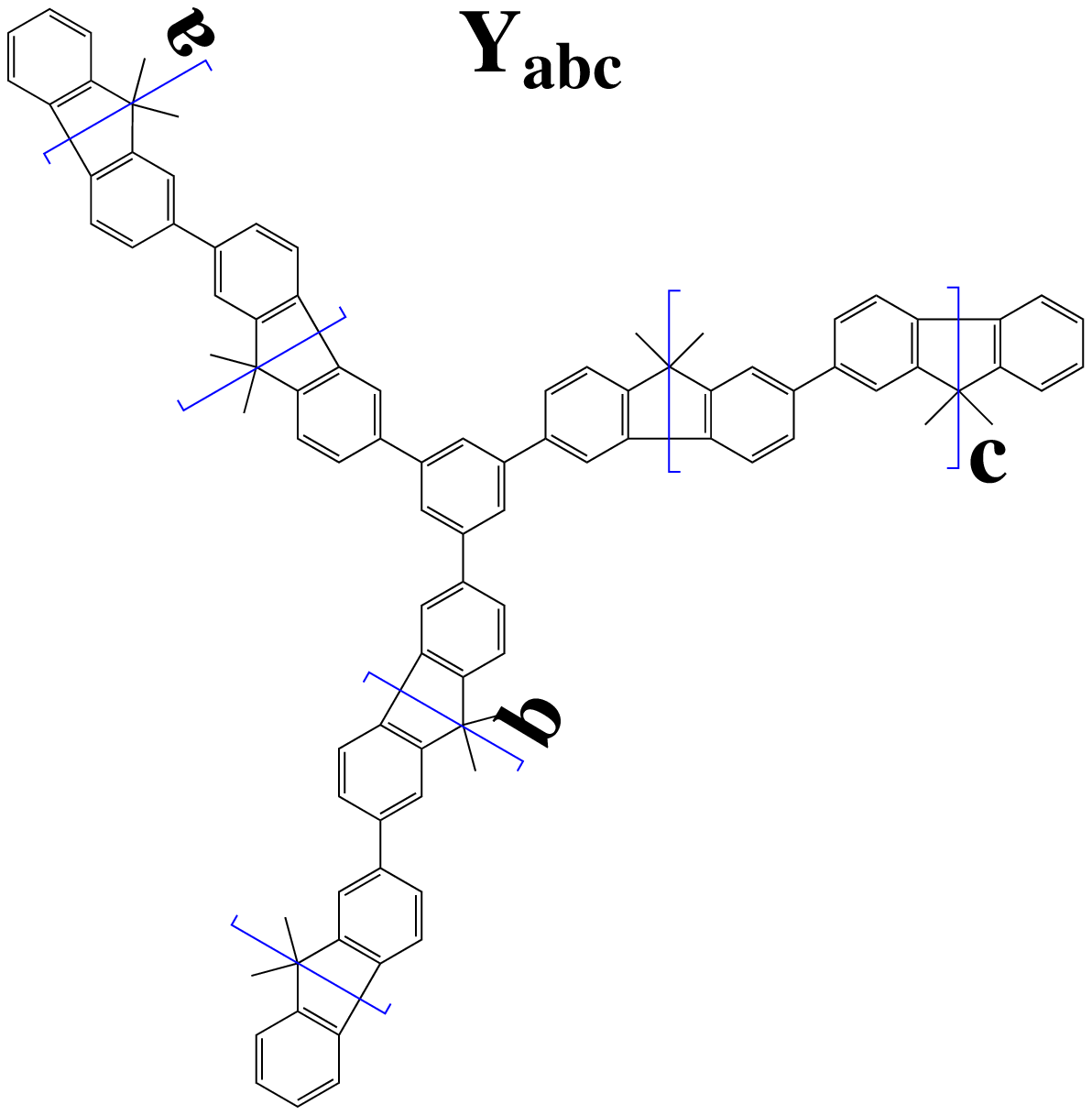}
    \includegraphics[width=0.35\textwidth]{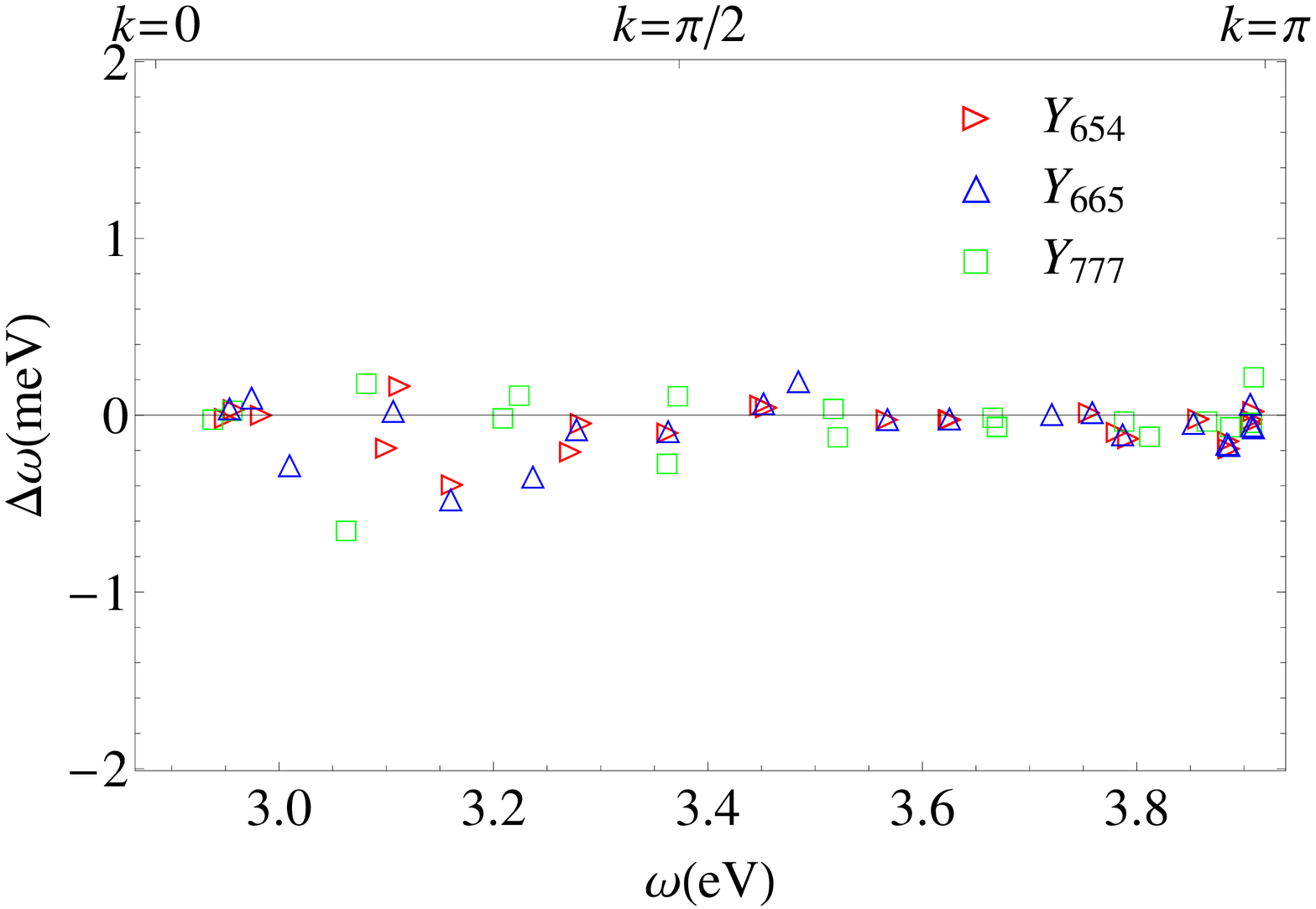}
  \end{center}
  \vspace{-0.6cm}
  \caption{(a) The Y-shape molecule Y$_{jmn}.$  (b) Differences between
excitation energies obtained by ES approach and the CEO method.
$\Delta\omega=\omega_{\rm ES}-\omega_{\rm CEO}$.}
 \label{energy}
\end{figure}

\begin{figure}[tpH]
  \begin{center}
    \includegraphics[width=0.35\textwidth]{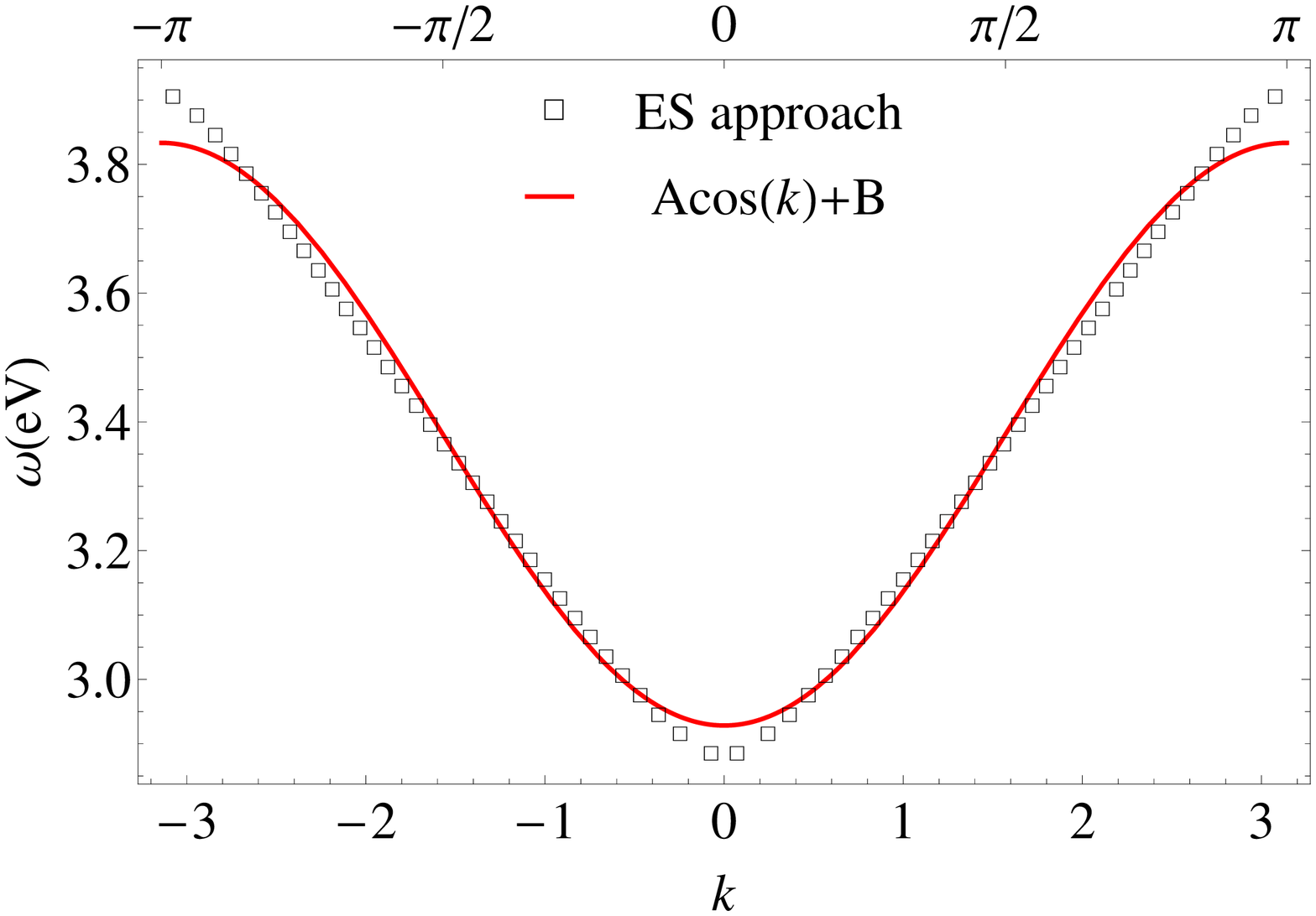}
    \includegraphics[width=0.35\textwidth]{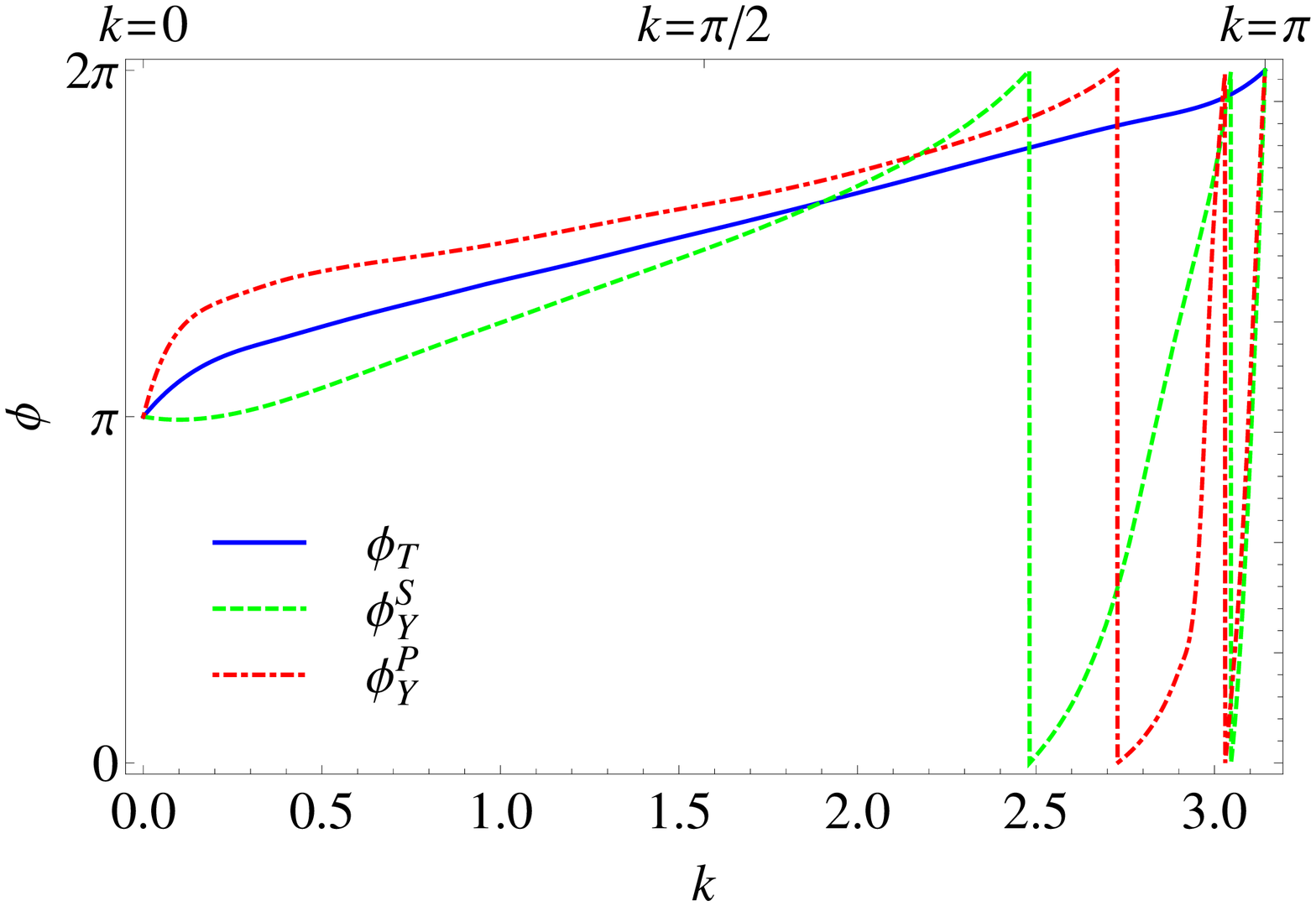}
    \vspace{-0.7cm}
  \end{center}
\caption{(a) The dispersion of the first exciton band fit
with a single cosine $A \cos(k) + B$, $A=-0.452$ and $B=3.381$. (b)
The scattering phases $\phi_{\rm T}$ of the terminals, and
$\phi_{\rm Y}^S$ and $\phi_{\rm Y}^P$ of the Y joint. Both $\phi_{\rm Y}^S$ and $\phi_{\rm Y}^P$ are shifted by
$-4\pi$. } \label{disp}
\end{figure}

We further illustrate the above theoretical models using quantum-chemical calculations of branched
polyfluorenes, a class of technologically important conjugated
polymers \cite{ScherfU:SempTr, BeckerS:Optple}. A Y-shaped family of molecules Y$_{abc}$ is
shown in Fig.~\ref{energy} (a), where $a$, $b$, and $c$ are the segment lengths
in repeat units. The ground state molecular geometries were optimized at the
AM1 level \cite{DewarMJS:AM1ngp} with Gaussian09 package \cite{g09}, followed by
computing the vertical excitation energies and transition density
matrices of up to 48 excited states using the Collective Electronic
Oscillator (CEO) method \cite{MukamelS:Elecco, TretiakS:Denmas}. We select
the first exciton band by checking the transition density matrices. Following our previous work \cite{ES-prl,ES3,ES-symmetric,ES-DA,ES-dipole} we further extracted the dispersion $\omega(k)$ (Fig.~\ref{disp} (a)) and scattering matrices $\Gamma (k)$ for $k\in[0,\pi]$. For the scattering matrix $\Gamma_{\rm T}$ of the terminal and  $\Gamma_{\rm Y}$ of Y joint, we have $\det(\Gamma_{\rm T})=e^{i\phi_{\rm T}}$ and $\det(\Gamma_{\rm Y})=e^{i\phi_{\rm Y}^S}e^{2i\phi_{\rm Y}^P}$ ,
where $\phi_{\rm Y}^S$ and $\phi_{\rm Y}^P$ are scattering phases of the $Y$-joint corresponding to a singlet and
2-fold degeneracy, respectively (see \cite{ES-symmetric} for details). The scattering parameters in Fig.~\ref{disp} allow for an accurate extraction of the excitation energies for an arbitrary Y-shaped molecules using the ES equation (\ref{ES-equations-vector}), see for example Fig.~\ref{energy} (b).

From the scattering phases (Fig.~\ref{disp} (b)), we extract the winding numbers $w(\Gamma_{\rm T})=1$ and $w(\phi_Y^S) = w(\phi_Y^P) = 5$, the latter following from each having 2 kinks (sharp $2\pi$ jumps) over half of the Brillouin zone, and a total shift of $2\pi$ over the full Brillouin
zone. Therefore, $w(\Gamma_{\rm Y})=w(\phi^S_{\rm Y})+2w(\phi^P_{\rm Y})=15$ and according to Eq.~(\ref{bound-explicit-intuit}), the number of solutions for the ES equation is
\begin{eqnarray}
\label{w-example}
\aligned
m=w(\tilde\Gamma)&=2(a+b+c)+3w(\Gamma_{\rm T})+w(\Gamma_{\rm Y})\\
&=2(a+b+c)+18
\endaligned
\end{eqnarray}
for the molecule Y$_{abc}$ shown in Fig.~\ref{energy} (a). There are a total of 6 solutions for $k=0$ and $k=\pi$, which do not correspond to physical states. Hence, the total number of states with excitation energies inside the exciton band is
\begin{eqnarray}
\label{N-example} N=(m-6)/2=a+b+c+6,
\end{eqnarray}
where the factor $1/2$ arises due to time reversal symmetry, in full accordance with Eq.~(\ref{m-to-N-intuitive}).

The result, presented in Eq.~(\ref{N-example}) can be obtained directly by using the general counting formula [Eq.~(\ref{exciton-bound-general})] with $l_{{\rm T}}= 1$, $l_{{\rm Y}}= 15$, $d_{0}= d_{\pi}= n_{{\rm s}}= 3$, or its generic counterpart [Eq.~(\ref{bound-symmetric})] with $Q_{{\rm T}}= 0$, $Q_{{\rm Y}}= 6$, where the vertex topological charges were computed using Eq.~(\ref{define-Q-a}).

\section{Conclusion}
\label{sec:Conclusion}

In summary, we have identified an integer-valued topological invariant $l_{a}$,
associated with an arbitrary scattering center $a$ of a branched molecular
structure (schematically represented by a graph), which is completely determined
by the corresponding scattering matrix $\Gamma_{a}(k)$. The integer $l_{a}$ was
motivated by a well known concept from algebraic topology, and has a simple integral form
\begin{eqnarray}
\label{define-l-a} l_{a}=\int_{-\pi}^{\pi}\frac{dk}{2\pi i}(\det \Gamma_{a}(k))^{-1}\frac{d}{dk}\det \Gamma_{a}(k).
\end{eqnarray}
We have reduced the problem of counting the number of solutions $m$ of the ES
equations to an intersection problem, and further extended topological
intersection theory to come up with a simple lower bound for $m$ in terms of
the segment lengths $L_{\alpha}$ and vertex winding numbers $l_{a}$
[Eq.~(\ref{bound-explicit-intuit})]. This allowed us to evaluate the total
number of electronic excitations $N$ within a spectral region of a single
exciton band [Eq.~(\ref{exciton-bound-general})].  Additionally, this permitted a simple and
easy to interpret formula [Eq.~(\ref{bound-symmetric})], that works for the generic
case. Furthermore, it does not contain the band-edge indices $d_{k}$, and counts the number of
excitations in terms of the vertex topological charges $Q_{a}= (l_{a}-
r_{a})/2$. We also demonstrated that both lower bounds become exact
for long enough linear segments. Quantum chemical
calculations of Y-shaped branched oligofluorenes provide direct illustration of
our theory. Being formulated in terms of dispersion relations and scattering
matrices, the presented counting expressions are very general. Therefore, this
approach can be applied to any quantum quasi-one-dimensional systems, or even
more generally to any wave phenomena in networks, e.g., optical communications.

The introduced concepts carry a number of possible implications on studies of
photo-physical phenomena in branched conjugated structures. The presented theory
provides an explicit link between the graph topology (connectivity) of the branched
system the interlining molecular electronic and optical properties defined by
the spectrum of electronic excitations: The excited-state electronic structure
in a finite molecule is a result of complex interplay between excitations in
linear segments, determined by the exciton band structure in infinite polymers
and electronic excitations in scattering centers. Chemical coupling between the
linear segments and scattering centers modifies the excitations in the latter,
resulting in bound and resonant states \cite{ES-analytical}. The resonant
states that fall inside the exciton band show up as resonances in the density
of states, or equivalently in sharp $2\pi$-rotations of the $\det\Gamma_{a}(k)$
in the narrow regions of $k$ around the resonance of $\omega(k)$ with a state
at the scattering center, referred to as kinks
\cite{{ES-tight-binding},ES-analytical} are much harder to analyze, compared to
bound states. Our counting result allows for a following interpretation: the
number $N_{{\rm res}}$ of resonant states in a finite structure is given by
just the sum $N_{{\rm res}}= \sum_{a}Q_{a}$ of the topological charges,
associated with the scattering centers. In particular, when the connectivity of
a conjugated structure is changed, with the scattering centers being kept the
same, the resonant states do not appear or disappear, but rather their
structure is modified. In the future we will study the dependence of the
resonant states structure on the graph topology by examining the position
dependent exciton density of states. 

Finally, we note that the possible increase in the number of states $N$ inside
the exciton band for shorter segments (i.e., in the case of inequality in
Eq.~(\ref{bound-symmetric})) can be explained by noting some of the bound
states get ``pushed'' inside the exciton band due to the effects of quantum
confinement of the exciton center of mass motion. We would like to emphasize
that the inequality that forms a lower bound for $N$ [Eq.~(\ref{bound-symmetric})],
which follows from topological considerations, reflects a physical/chemical
trend: Effects of quantum confinement of the exciton center of mass motion are
sometimes turning the bound states into the resonant counterparts and not the
other way around.




\begin{acknowledgments}

This work is supported by the National Science Foundation under Grant No. CHE-1111350, and U.S. Department of Energy and Los Alamos LDRD funds. Los Alamos National Laboratory is operated by Los Alamos National Security, LLC, for the National Nuclear Security Administration of the U.S. Department of Energy under Contract No. DE-AC52-05NA25396. We acknowledge support of Center for Integrated Nanotechnology (CINT). We would also like to thank John Klein for helpful discussions regarding the index theorem.
\end{acknowledgments}

\appendix

\section{Topological Properties of the Unitary Groups}
\label{app:topological properties}

In this appendix, we describe some relevant topological
properties of the unitary groups $U(n)$ in a simple and self-consistent
fashion.

First, taking the determinant of a unitary matrix defines a map ${\rm
det}:U(n)\to U(1)$. We can define a map $s:U(1)\to U(n)$, referred to as a
section, by defining $s(\lambda)$ to be a unitary matrix with the eigenvalues
$(\lambda,1,\ldots,1)$. Note that ${\rm det}\circ s={\rm id}$. A choice of a
section defines an isomorphism (of topological spaces, as well as smooth
manifolds, but not groups) $u:U(n)\to U(1)\times SU(n)$, given by
$u(g)=(\det{g},g(s(\det{g}))^{-1})$. We are interested in topological
invariants of maps $S^{1}\to U(n)$, or more precisely the homotopy
equivalence classes of such maps. Since $U(n)\cong U(1)\times
SU(n)$, a map $S^{1}\to U(n)$ is described by a pair of maps $S^{1}\to U(1)$
and $S^{1}\to SU(n)$. If $[X,Y]$ denotes the set of homotopy
classes of maps $X\to Y$, then we have $[S^{1},U(n)]\cong
[S^{1},U(1)]\times [S^{1},SU(n)]$. Using the fact that $[S^{1},SU(n)]$ consists of a single point,
each $1$-dimensional cycle in $SU(n)$ is contractible (meaning
topologically equivalent to the constant map).
As a topological space $U(1)\cong S^{1}$, so that $[S^{1},U(1)]\cong
[S^{1},S^{1}]\cong \mathbb{Z}$, with the homotopy class of a map
given by its winding number. Therefore $[S^{1},U(n)]\cong
\mathbb{Z}$, i.e., maps from a circle to a unitary group have exactly one
topological invariant that takes values in the integers $\mathbb{Z}$.
This also implies that the determinant map ${\rm det}:U(n)\to U(1)$
generates an isomorphism $[S^{1},U(n)]\to [S^{1},U(1)]\cong \mathbb{Z}$.
Formulated in a simpler way, to identify the topological invariant of a map
$f:S^{1}\to U(n)$ one needs just to compute the winding number $w({\rm
det}(f))$, where ${\rm det}(f)={\rm det}\circ f$. Therefore, we refer to the
topological invariant $w(f)=w({\rm det}(f))$ of a map $f:S^{1}\to U(n)$ as
its winding number.

In proving the index theorem (see a sketch in
Section~\ref{sec:sketch-proof-index}), we will make use of the fact that
the space $U(n)\setminus D_{1}U(n)$ is contractible to a point. We reiterate
that $D_{1}U(n)\subset U(n)$ is a closed subspace of unitary matrices that have
at least one unit eigenvalue. In particular, if a map $f:S^{1}\to U(n)$ misses
$D_{1}U(n)$, i.e., can be represented as a composition $S^{1}\to U(n)\setminus
D_{1}U(n)\to U(n)$ (where the second map is the inclusion), then $w(f)=0$. The
contractibility can be demonstrated by presenting a continuous deformation
(homotopy) of the identity map ${\rm id}:U(n)\setminus D_{1}U(n)\to
U(n)\setminus D_{1}U(n)$ to a constant map, the latter mapping the whole domain
to a single point. The spectra of matrices $x\in U(n)\setminus D_{1}U(n)$ are
contained in an open interval $(0,2\pi)$ in the sense that the eigenvalues are
given by $\lambda_{s}=e^{i\varphi_{s}}$ with $\varphi_{s}\in (0,2\pi)$.
Consider an obvious contraction of the interval to its center, given by the
point $\pi\in (0,2\pi)$. The contraction/deformation of the interval defines a
corresponding contraction/deformation of the spectrum of any matrix $x\in
U(n)\setminus D_{1}U(n)$. The described above deformation of the spectrum
provides a deformation of the corresponding matrix: we deform the spectrum
keeping the eigenspaces, related to the eigenvalues unchanged. This describes a
contraction of $U(n)\setminus D_{1}U(n)$ to the point $-{\rm id}\in
U(n)\setminus D_{1}U(n)$.

\end{document}